\documentclass{easychair}

\usepackage[utf8x]{inputenc}
\usepackage{amsfonts}
\usepackage{hyperref}

\usepackage{graphicx}

\usepackage{doc}
\usepackage{tikz}
\usetikzlibrary{arrows.meta}
\usetikzlibrary{shapes.multipart}
\usepackage{amsfonts}
\usepackage[linesnumbered]{algorithm2e}

\title{A Note on the Performance of Algorithms for Solving Linear
  Diophantine Equations in the Naturals}

\author{
Valeriu Motroi\inst{1}
\and
{\c S}tefan Ciob{\^ a}c{\u a}\inst{2}
}

\institute{
  Alexandru Ioan Cuza University
  Ia{\c s}i, Romania\\
  \email{\{motroival, stefan.ciobaca\}@gmail.com}
}

\authorrunning{Motroi and Ciob{\^ a}c{\u a}}

\titlerunning{A Note on the Performance of Algorithms for Solving LDE in the Naturals}

\begin{document}

\maketitle

\begin{abstract}
  We implement four algorithms for solving linear Diophantine
  equations in the naturals: a lexicographic enumeration algorithm, a
  completion procedure, a graph-based algorithm, and the Slopes
  algorithm. As already known, the lexicographic enumeration algorithm
  and the completion procedure are slower than the other two
  algorithms. We compare in more detail the graph-based algorithm and
  the Slopes algorithm. In contrast to previous comparisons, our work
  suggests that they are equally fast on small inputs, but the
  graph-based algorithm gets much faster as the input grows. We
  conclude that implementations of AC-unification algorithms should
  use the graph-based algorithm for maximum efficiency.

\end{abstract}

\maketitle
\section{Introduction}\label{sec:intro}
Solving linear Diophantine equations in the naturals is at the core of
AC-unification algorithms. AC-unification reduces to top-most
unification problems of the form
\[ f^*(u_1, \ldots, u_l) = f^*(v_1, \ldots, v_k), \] where
\( u_1, \ldots, u_l, v_1, \ldots, v_k \) are variables (possibly with
repetitions) and \( f^* \) is a variadic symbol corresponding to some
AC-symbol \( f \). Solving such a top-most AC-unification problem
reduces to solving the Diophantine equation  

\begin{equation}\label{eq:lindioph}
  a_1x_1+a_2x_2+\cdots+a_nx_n=b_1x_1+b_2x_2+\cdots+b_nx_n, 
\end{equation}

\noindent where \( x_i \) (\( 1 \leq i \leq n \)) are the unknowns
(taking values in the set of naturals) and \( a_i, b_j \)
(\( 1 \leq i \leq l \), \( 1 \leq j \leq k\)) are the multiplicities
of the corresponding variable among \( u_1, \ldots, u_l \) and
\( v_1, \ldots, v_k \), respectively. For more details, see the survey
of Baader and Snyder on
unification~\cite{DBLP:books/el/RV01/BaaderS01}.

Therefore, algorithms for computing AC-unifiers make intensive use of a
linear Diophantine equation (LDE) solver. We compare four algorithms
for LDE solving. Our results suggest that the graph-based algorithm is
the fastest on modern computers, in contrast to previous benchmarks,
which are older. We conclude that implementations of AC-unification
should consider switching to the graph-based algorithm.

\section{Algorithms}\label{sec:algos}

In this section we briefly describe the known algorithms for solving
Equation~\ref{eq:lindioph}.

\subsection{The Lexicographic Enumeration Algorithm}

In this subsection we describe the simplest way of solving
Equation~\ref{eq:lindioph}. We notice that a linear Diophantine
equation with natural solutions can have an infinite number of
solutions. We generally do not need all of them, but just need a
complete set of minimal solutions. A solution
$S_1=(x_1, x_2, \ldots, x_n)$ is not minimal if there exists another
solution $S_2=(x'_1, x'_2, \ldots, x'_n)$ such that for all $i$,
$S_{1, i} \geq S_{2,i}$ and $S_1 \neq S_2$. The set of minimal
solutions forms a basis. The lexicographic
algorithm~\cite{HUET1978144} lexicographically enumerates all
solutions and saves only the minimal ones. However, we can not
enumerate infinitely many solutions; we should have a bound for
$x_{a,i}$ and $x_{b,i}$, where the vectors $x_a$ and $x_b$ form a
solution of Equation~\ref{eq:lindioph}. Huet~\cite{HUET1978144} points
out that, for a minimal solution, the unknowns $x_{a,i}$ should be not
greater than $\max(b)$ and the unknowns $x_{b,i}$ should not be greater
than $\max(a)$, where $a$ and $b$ are the coefficients in
Equation~\ref{eq:lindioph}. Lambert~\cite{lambert1987borne} gives the
stronger bounds $\sum_i x_{a,i} \leq \max(b)$ and
$\sum_i x_{b,i} \leq \max(a)$. Moreover, we do not need to enumerate
the possible values of all $N$ unknowns, it is enough to enumerate
$N-1$; the last unknown can be found by solving a simple equation. We
can develop this idea further. What happens if we enumerate $N-2$
variables? We get an equation of form $ax+by=c$ and we can solve it
using the extended Euclidian algorithm. With two types of bounds and
two types of \emph{optimizations} we get 4 similar algorithms that
solve Equation~\ref{eq:lindioph}. The slowest part of this algorithm
is checking if a new solution is minimal or not, because checking
minimality requires comparing the new solution with the other already
generated minimal solutions. For implementation simplicity we rewrite
Equation~\ref{eq:lindioph} as:
\begin{equation}\label{eq:lindioph2}
a_1x_1+a_2x_2+\ldots+a_nx_n - b_1x_1-b_2x_2-\ldots-b_nx_n=0
\end{equation}
If we let $w_i = a_i-b_i$ then Equation~\ref{eq:lindioph2} can be written as:
\begin{equation}\label{eq:lindioph3}
w_1x_1+w_2x_2+\ldots+w_nx_n=0
\end{equation}
We work with Equation~\ref{eq:lindioph3}, because it is closer to the
implementation.  In Figure~\ref{fig:lexalg} we provide a very generic
implementation for the lexicographic algorithm. The algorithm
implements a standard backtracking procedure, with the parameter $p$
denoting the current unknown. The test \textit{$p$ is last} allows to
implement the optimizations described above (stop the enumeration at
$n-1$ unknowns or at $n-2$ unknowns).

\begin{figure}
\begin{algorithm}[H]
\DontPrintSemicolon
\SetKwFunction{FLexAlg}{LexAlg}
\SetKwProg{Fn}{Function}{:}{}
\SetKwFor{For}{for (}{)}{}
\Fn{\FLexAlg{p}} {
  \eIf{p is last} {
    sol := Solve the remaining equation without enumeration\;
    
    \uIf{sol is Minimal} {
      addSolution(sol)\;
    }
  } {
    \For{$i = 0;\ i < bound;\ i = i + 1$} {
      set sol[p] = i\;
      
      LexAlg(p + 1)\;
    }
  }
}
\end{algorithm}
\caption{\label{fig:lexalg}The Lexicographic Enumeration Algorithm (generic version).}
\end{figure}

\subsection{The Completion Procedure}\label{sec:CompProc}
Another way to solve Equation~\ref{eq:lindioph} is to compute all
minimal solutions by a \emph{completion procedure}. Such an algorithm
is due to Fortenbacher~\cite{fortenbacher1983algebraische}, with an
optimization by Guckenbiehl and
Herold~\cite{guckenbiehl1985solving}. For some
$x=(x_1, x_2, \ldots, x_n)$, we denote by $d(x)$ the result of the
expression $d(x)=w_1x_1+w_2x_2+\ldots+w_nx_n$, which we call the defect
of Equation~\ref{eq:lindioph3}. A \emph{proposal}
$p = (x_1, x_2, \cdots, x_n)$ is characterized by
$-\max b \leq d(p) \leq \max a$. A solution is a proposal \( p \) that
has $d(p) = 0$. The algorithm starts with a set of proposals. At each
\emph{completion step}, it updates every proposal $p$ in the following
way: \( \bullet \) if its defect is less than zero, then it increments
$x_i$ by $1$ for some index $i$ with $w_i > 0$; \( \bullet \)
otherwise (if its defect is positive) it increments $x_i$ by $1$ for
some $i$ with $w_i < 0$. If the result has defect zero then a
\emph{minimal solution} was found. If a proposal is not minimal then
it is discarded, because we can not obtain a \emph{minimal solution}
from a \emph{non-minimal proposal}. In such a way only minimal
solutions are computed and this is an advantage over
\emph{Lexicographic Algorithm}. However, a solution may be computed
several times and we still have to test proposals for
minimality. Guckenbiehl and Herold~\cite{guckenbiehl1985solving}
describe a way to avoid computation of the same solution several
times. To do that we need to select one unique computation for each
solution. That is done by following one rule: a proposal with negative
(positive) defect must not be incremented at a position $i$ (position
$j$) if there exists a $k > i$ with $w_k > 0$ ($k > j$ with
$w_k < 0$). We present this version of the completion procedure in
Figure~\ref{fig:compalg}.

\begin{figure}
\begin{algorithm}[H]
\DontPrintSemicolon
\SetKwFunction{CompAlg}{CompAlg}
\SetKwProg{Fn}{Function}{:}{}
\SetKwFor{For}{for (}{)}{}

\Fn{\CompAlg{w}} {
  // init the proposal set\;
  pSet = empty set of proposals\;
  \For{$i = 0;\ i < n;\ i = i + 1$} {
    p = new proposal\;
    p[i] = 1\;
    pSet.add(p)\;
  }
  
  \While{pSet is not empty} {
    // completion step\;
    pSetNew = a new proposal set\;
    \For{$p\ in\ pSet$} {
      \For {$i = n - 1;\ i \geq 0;\ i = i - 1$} {
        \uIf{$(d(p) < 0\ and\ w_i < 0)\ or\ (d(p) > 0\ and\ w_i > 0)$} {
          continue\;
        }
        auxProposal = copyOf(p)\;
        auxProposal[i] = auxProposal[i] + 1\;
        \eIf{auxProposal has defect zero} {
          addSolution(auxProposal)\;
        } {
          \uIf{auxProposal is minimal} {
            pSetNew.add(auxProposal)\;
          }
        } 
        // avoiding multiple computations of the same solution\;
        \uIf{$p[i] > 0$} {
          break\;
        }
      }
    }
    pSet = pSetNew\;
  }
}

\end{algorithm}
\caption{\label{fig:compalg}The Completion Procedure described by
  Fortenbacher~\cite{fortenbacher1983algebraische}, with an
  optimization by Guckenbiehl and
  Herold~\cite{guckenbiehl1985solving}.}
\end{figure}

\subsection{The Graph Algorithm}\label{sec:GrAlg}
Clausen and Fortenbacher~\cite{clausen1989efficient} described a
further optimization of the completion procedure. We call the
resulting algorithm the \emph{Graph Algorithm}. In order to avoid all
additions and subtractions, they represent Equation~\ref{eq:lindioph3}
as a graph. The graph representation of a linear Diophantine equation
is a labelled digraph with the set
$\{d \in \mathbb{Z} | - \max b \leq d \leq \max a\}$ representing the
nodes and set $\{d \rightarrow_{w_i} d + w_i | i \leq n\}$
representing labelled edges. In other words, the nodes are the defect
of any proposal and an edge $d \rightarrow_{w_i} d + w_i$ corresponds
to incrementing a proposal at position $i$. A solution in this graph
is a walk that begins in zero and ends in zero. An advantage over the
\emph{completion procedure} is that the minimality check of a solution
is also transformed into a graph problem. In this graph, a walk that
corresponds to the solution $s = (s_1, s_2, \ldots, s_n)$ is
non-minimal if there is another walk $z = (z_1, z_2, \ldots, z_n)$
that is shorter than $s$ and also is bounded by $s$, in other words
$z_i \leq s_i$ for all $i$. A detailed implementation of this
algorithm written in Pascal is provided by Clausen and
Fortenbacher~\cite{clausen1989efficient}.

\subsection{The Slopes Algorithm}\label{sec:SlopesAlg}
The Slopes Algorithm described by Filgueiras and Tom{\'a}s~\cite{filgueiras1995fast} is an optimization of the lexicographic algorithm. The enumeration is performed \emph{for all but three} of the unknowns and an equation of the following form:
\begin{equation}\label{eq:lde3unk}
  ax = by + cz + v,\ a,b,c,x,y,z \in \mathbb{N},\ v \in \mathbb{Z}
\end{equation}
is solved. Filgueiras and Tom{\'a}s~\cite{filgueiras1995fast} describe
a way of finding directly all minimal solutions for
Equation~\ref{eq:lde3unk}. The idea is that, if minimal solutions of
Equation~\ref{eq:lde3unk} are ordered with $z$ strictly increasing,
then both the solution with the smallest $z$ and the difference
between consecutive solutions can be computed
algebraically. Geometrically, this can be seen as a Pareto frontier of
all solutions projected onto the \emph{YZ-plane} if $v \geq 0$ and is
a polygonal line when $v < 0$. We can project solutions to 2D space
because it is well known that each solution of
Equation~\ref{eq:lde3unk} verifies the congruence:
\begin{equation}\label{eq:lde3unk2dspace}
by + cz \equiv -v\ \ \ (mod\ a).
\end{equation}
And reciprocally, that each solution of
Equation~\ref{eq:lde3unk2dspace} corresponds to some integral solution
of Equation~\ref{eq:lde3unk}. In Figure~\ref{fig:slopesalg} we present
the implementation of Slopes algorithm for solving $ax + by + cz = 0$.

\begin{figure}
\begin{algorithm}[H]
\SetKwFunction{Slopes}{Slopes}
\SetKwProg{Fn}{Function}{:}{}
\SetKwFor{For}{for (}{)}{}
\Fn{\Slopes{a, b, c}} {
  gb=gcd(a,b); gc=gdc(a,c); G=gcd(gb,c)\;
  ymax=a/gb; zmax=a/gc\;
  dz=gb/G; dy=(c*multiplier(b,a)/G) mod ymax\;
  y=ymax-dy; z=dz\;
  Solutions=\{(b/gb, ymax, 0), (c/gc, 0, zmax), ((b*y+c*z)/a, y, z)\}\;
  \While{dy $>$ 0} {
    \While{y $>$ dy} {
      y=y-dy; z=z+dz\;
      Solutions.add(((b*y)+c*z)/a,y,z))\;
    }
    f=dy/y; dy=dy mod y; dz=f*z+dz\;
  }
  \Return Solutions\;
}
// multiplier(a, b) is an integer $m_b$ such that $gcd(a,b)=m_a*a + m_b*b$
\end{algorithm}
\caption{\label{fig:slopesalg}The Slopes algorithm of Filgueiras and
  Tom{\'a}s for the equation $ax = by + cz$.}
\end{figure}

\section{Methodology}\label{sec:Meth}

We have implemented all algorithms in C++. The first two algorithms
are clearly slower than the last two. Therefore, we made a more
detailed comparison between the Graph and Slopes algorithms,
reproducing the comparison by Filgueiras and
Tom{\'a}s~\cite{filgueiras1995fast}. As our implementation of the
Slopes algorithm is somewhat slower than the well known and optimized
C implementation~\cite{Slopes}, we use the later for the
comparison. All of our code, including instructions for reproducing
our results (Figures 1-4), are available at
\url{https://github.com/Djok216/LDEAlgsComparison}.

To measure the running time we use the python \emph{subprocess} and
\emph{time} libraries. The first one is used for spawning the
executables of the algorithms and the later for measuring running time
using \emph{perf\_counter}. We set a timeout of $10$ minutes for the
spawned processes. The tests are generated using \emph{random.randint}
and the left side is sorted in decreasing order and the right side in
increasing order, because in most cases this ordering speeds up the
Slopes algorithm as explained by Filgueiras and
Tom{\'a}s~\cite{filgueiras1995fast}.

The tests are divided in $160$ classes determined by
$N \in \{ 1,2,3,4 \}$ - the number of unknowns on the left hand side,
$M \in \{ 2, 3, 4, 5, 6, 7, 8, 9 \}$ - the number of unknowns on the
right hand side such that $N \leq M$ and
$\emph{MaxValue} \in \{ 2, 3, 5, 13, 29, 39, 107, 503, 1021 \}$- the
maximum coefficient of any unknown. We manually set \emph{MaxValue} as
part of the coefficients on the right hand side, because there are
always more unknowns on the right hand side than left hand side
($N \leq M$). Every class contains $10$ different tests generated
randomly. We use a fixed seed for reproducibility purposes.

We calculate the running time after running the same test with the
same algorithm $5$ times. The running time for that algorithm on that
specific test is considered to be the arithmetic mean of $3$ out of
$5$ remaining values after removing the smallest and the biggest
value. The exception is when an algorithm runs for more than $15$
seconds. In this case we stop running the same test and calculate the
arithmetic mean of running times available at the moment. For example,
if an algorithm runs two times in $14.9$ seconds and the third time in
$15.2$, then we stop running this test after third run and the time is
considered to be $(2 * 14.9 + 15.2) / 3 = 15$ seconds. We also set a
timeout of $10$ minutes, after which we automatically stop the
algorithm.

For every test in a given class, we add $1$ point to the algorithm
taking the least time and $0$ points to the other. In case of a tie,
we add $0.5$ to both algorithms. Therefore, a score of $6:4$ would
mean that the first algorithm performed better $6$ times, while the
second algorithm $4$ times out of the $10$ tests for a given class.

We consider an algorithm to win a particular class if it scores at
least $8$ points. The definition of a win is justified statistically
by Filgueiras and Tom{\'a}s~\cite{filgueiras1995fast}.

For compiling the code we use \emph{GCC 5.4}. Below are the commands
used to compile the programs:
\begin{verbatim}
gcc -static slopesV7i.c -std=c11 -O3 -o slopesV7i
g++ -static -lm -s -x c++ -std=c++17 -O3 -o graph graph.cpp 
\end{verbatim}
\noindent We run the benchmark on an Intel Xeon machine with two
processors and 24 hardware threads (12 physical cores) with a clock
speed of 2.67 GHz.

We repeat the measurements made by Filgueiras and
Tom{\'a}s~\cite{filgueiras1995fast}, but we also compare the two
algorithms using an epsilon of $0.01$. By this we mean that the
algorithms are considered equally fast if their running times differ
by a value smaller than $0.01$ seconds. Moreover, we also compute the
overall time spent in every class of tests.

\section{Discussion}\label{sec:Dis}

Figure~\ref{fig1} contains a summary of the results. The Slopes
algorithm wins $103$ classes out of $160$ and Graph algorithm wins $7$
classes. Filgueiras and Tom{\'a}s~\cite{filgueiras1995fast} find that
Slopes wins $88$ classes and Graph wins $33$ classes. These results
suggest that the Slopes algorithm is faster than the Graph algorithm.

We redo the same comparison, but this time we consider the two
algorithms to be equal if their running time differs by at most
$\epsilon = 0.01$ seconds. The results are summarized in
Figure~\ref{fig2}. Each algorithm now has $6$ wins and for most of the
classes there is a tie. This means that the algorithms are quasi-equal
in efficiency.

Going further, we analyze the \emph{total time} spent in each test
class by each algorithm. The results are summarized in
Figure~\ref{fig4}. We see that in classes where the Slopes algorithm
wins, the difference is very small. However, in the classes in which
the Graph algorithm wins, the difference is huge. The total time spent
in all $160$ classes for the Slopes algorithm is $4284.81$ seconds and
$724$ seconds for the Graph algorithm. The counts ($4284.81$ seconds,
$724$ seconds) should be interpreted taking into account that they
contain $1$ timeout of $10$ minutes for Graph and $4$ timeouts of $10$
minutes for Slopes, as summarized in Figure~\ref{fig3}.

Based on our results, we conclude that the Graph algorithm is
significantly faster than Slopes for bigger instances and roughly as
fast for small instances.

\emph{Practical relevance of our benchmark.} In most cases, the
bottleneck in AC(U)-unification algorithms is combining the solutions
to the linear Diophantine equations themselves. However, there are
AC(U)-unification problems where solving Equation~\ref{eq:lindioph} is
the slow part. An example is an ACU-unification problem with a single
AC-function $f$ and 8 different variables, which can be constructed
based on Equation~\ref{eqToSolve}:
\begin{equation}\label{eqToSolve}
  104x_1 + 167x_2 = 165x_3 + 154x_4 + 148x_5 + 159x_6 + 174x_7 + 150x_8.
\end{equation}
The ACU-unification problem is the following:
\begin{equation}\label{eqToSolveProb}
  f_{104}(u_1) + f_{167}(u_2) =_? f_{165}(u_3) + f_{154}(u_4) + f_{148}(u_5) + f_{159}(u_6) + f_{174}(u_7) + f_{150}(u_8),
\end{equation}
\noindent where $f_k(v) = f(v, v, \ldots, v)$ (such that $v$ has $k$
occurrences).

Equation~\ref{eqToSolve} has a basis of size $5510$. Finding the basis
is significantly slower than combining its solutions and creating the
ACU-unifier. On the same hardware as described in
Section~\ref{sec:Meth}, the Graph algorithm takes about $0.6$ seconds
to solve the linear Diophantine equation above, while combining the
solutions into an ACU-unifier takes $0.15$ seconds. To compute the
ACU-unifier, we use the algorithm presented by Baader and Snyder in
their survey on unification~\cite{DBLP:books/el/RV01/BaaderS01}.
Therefore, at least on some AC-unification problems, solving LDEs
dominates the running time.

\emph{Conclusion.} Implementations of AC unification should therefore
consider using the Graph algorithm, or choosing between Graph and
Slopes, depending on problem size.

\bibliography{comparison}
\bibliographystyle{plain}

\appendix

\begin{figure}
\begin{center}
\begin{tabular}{c@{\hspace{0.5cm}}c@{\hspace{0.5cm}}c@{\hspace{0.5cm}}c@{\hspace{0.5cm}}c@{\hspace{0.5cm}}c@{\hspace{0.5cm}}c@{\hspace{0.5cm}}c@{\hspace{0.5cm}}c@{\hspace{0.5cm}}c@{\hspace{0.5cm}}c}
\hline
& A & 2 & 3 & 5 & 13 & 29 & 39 & 107 & 503 & 1021 \\
N & M \\
\hline
1 & 2 & 2:8  & 2:8  & 2:8  & 5:5  & 2:8  & 4:6  & 4:6  & 1:9  & 0:10 \\
1 & 3 & 2:8  & 2:8  & 3:7  & 2:8  & 1:9  & 3:7  & 1:9  & 0:10  & 0:10 \\
1 & 4 & 3:7  & 3:7  & 2:8  & 2:8  & 1:9  & 0:10  & 0:10  & 0:10  & 0:10 \\
1 & 5 & 4:6  & 2:8  & 2:8  & 2:8  & 0:10  & 3:7  & 1:9  & 0:10  & 0:10 \\
1 & 6 & 3:7  & 3:7  & 2:8  & 1:9  & 0:10  & 1:9  & 0:10  & 0:10  & \\
1 & 7 & 4:6  & 4:6  & 3:7  & 2:8  & 1:9  & 0:10  & 0:10  & 0:10  & \\
1 & 8 & 1:9  & 2:8  & 3:7  & 0:10  & 0:10  & 0:10  & 0:10  &  & \\
1 & 9 & 4:6  & 2:8  & 1:9  & 1:9  & 2:8  & 1:9  &  &  & \\
\hline
2 & 2 & 3:7  & 1:9  & 1:9  & 2:8  & 1:9  & 0:10  & 1:9  & 0:10  & 0:10 \\
2 & 3 & 3:7  & 2:8  & 2:8  & 1:9  & 0:10  & 2:8  & 4:6  & 10:0  & 7:3 \\
2 & 4 & 2:8  & 3:7  & 3:7  & 0:10  & 1:9  & 3:7  & 6:4  & 9:1  & 9:1 \\
2 & 5 & 5:5  & 5:5  & 3:7  & 3:7  & 3:7  & 6:4  & 9:1  & 9:1  & \\
2 & 6 & 2:8  & 2:8  & 0:10  & 0:10  & 2:8  & 5:5  & 6:4  &  & \\
2 & 7 & 4:6  & 1:9  & 3:7  & 0:10  & 1:9  & 6:4  &  &  & \\
2 & 8 & 4:6  & 4:6  & 1:9  & 2:8  & 4:6  & 9:1  &  &  & \\
\hline
3 & 3 & 2:8  & 0:10  & 2:8  & 0:10  & 0:10  & 2:8  & 2:8  & 7:3  & \\
3 & 4 & 2:8  & 3:7  & 3:7  & 2:8  & 0:10  & 2:8  & 4:6  &  & \\
3 & 5 & 4:6  & 2:8  & 2:8  & 1:9  & 2:8  & 5:5  & 9:1  &  & \\
3 & 6 & 2:8  & 2:8  & 1:9  & 0:10  & 3:7  & 4:6  &  &  & \\
\hline
4 & 4 & 4:6  & 2:8  & 1:9  & 0:10  & 2:8  & 0:10  & 5:5  &  & \\
4 & 5 & 3:7  & 2:8  & 1:9  & 0:10  & 2:8  & 5:5  &  &  & \\

\hline
\end{tabular}
\caption{
  \label{fig1}Comparison of Graph and Slopes algorithms. For each of
  the $160$ test classes characterized by $N, M$ and $A$
  (\textit{MaxValue}), we count the number of times Graph is faster
  versus the number of times Slopes is faster out of $10$ tests for
  each class.}
\end{center}
\end{figure}

\begin{figure}
\begin{center}
\begin{tabular}{c@{\hspace{0.5cm}}c@{\hspace{0.5cm}}c@{\hspace{0.5cm}}c@{\hspace{0.5cm}}c@{\hspace{0.5cm}}c@{\hspace{0.5cm}}c@{\hspace{0.5cm}}c@{\hspace{0.5cm}}c@{\hspace{0.5cm}}c@{\hspace{0.5cm}}c}
\hline
& A & 2 & 3 & 5 & 13 & 29 & 39 & 107 & 503 & 1021 \\
N & M \\
\hline
1 & 2 & 5:5  & 5:5  & 5:5  & 5:5  & 5:5  & 5:5  & 5:5  & 5:5  & 4:5 \\
1 & 3 & 5:5  & 5:5  & 5:5  & 5:5  & 5:5  & 5:5  & 5:5  & 4:5  & 4:5 \\
1 & 4 & 5:5  & 5:5  & 5:5  & 5:5  & 5:5  & 5:5  & 5:5  & 2:7  & 1:9 \\
1 & 5 & 5:5  & 5:5  & 5:5  & 5:5  & 5:5  & 5:5  & 4:5  & 0:9  & 0:10 \\
1 & 6 & 5:5  & 5:5  & 5:5  & 5:5  & 5:5  & 5:5  & 2:7  & 1:9  & \\
1 & 7 & 5:5  & 5:5  & 5:5  & 5:5  & 5:5  & 4:5  & 2:7  & 0:9  & \\
1 & 8 & 5:5  & 5:5  & 5:5  & 5:5  & 4:6  & 4:5  & 2:8  &  & \\
1 & 9 & 5:5  & 5:5  & 5:5  & 5:5  & 4:6  & 3:7  &  &  & \\
\hline
2 & 2 & 5:5  & 5:5  & 5:5  & 5:5  & 5:5  & 5:5  & 5:5  & 4:5  & 3:6 \\
2 & 3 & 5:5  & 5:5  & 5:5  & 5:5  & 5:5  & 5:5  & 5:5  & 9:0  & 6:3 \\
2 & 4 & 5:5  & 5:5  & 5:5  & 5:5  & 5:5  & 5:5  & 7:3  & 9:1  & 9:1 \\
2 & 5 & 5:5  & 5:5  & 5:5  & 5:5  & 5:5  & 5:4  & 9:1  & 9:1  & \\
2 & 6 & 5:5  & 5:5  & 5:5  & 5:5  & 5:5  & 5:5  & 6:3  &  & \\
2 & 7 & 5:5  & 5:5  & 5:5  & 5:5  & 5:5  & 5:5  &  &  & \\
2 & 8 & 5:5  & 5:5  & 5:5  & 5:5  & 6:4  & 6:3  &  &  & \\
\hline
3 & 3 & 5:5  & 5:5  & 5:5  & 5:5  & 5:5  & 5:5  & 4:5  & 7:3  & \\
3 & 4 & 5:5  & 5:5  & 5:5  & 5:5  & 5:5  & 5:5  & 3:6  &  & \\
3 & 5 & 5:5  & 5:5  & 5:5  & 5:5  & 5:5  & 5:4  & 9:0  &  & \\
3 & 6 & 5:5  & 5:5  & 5:5  & 5:5  & 5:4  & 4:5  &  &  & \\
\hline
4 & 4 & 5:5  & 5:5  & 5:5  & 5:5  & 5:4  & 4:6  & 5:5  &  & \\
4 & 5 & 5:5  & 5:5  & 5:5  & 5:5  & 5:5  & 6:3  &  &  & \\

\hline
\end{tabular}
\caption{
  \label{fig2}Comparison of the Graph and Slopes algorithms. Same as
  the previous figure, but the algorithms are considered tied on tests
  on which their running times differ by at most $0.01$ seconds.}
\end{center}
\end{figure}

\begin{figure}
\begin{center}
\begin{tabular}{c@{\hspace{0.5cm}}c@{\hspace{0.5cm}}c@{\hspace{0.5cm}}c@{\hspace{0.5cm}}c@{\hspace{0.5cm}}c@{\hspace{0.5cm}}c@{\hspace{0.5cm}}c@{\hspace{0.5cm}}c@{\hspace{0.5cm}}c@{\hspace{0.5cm}}c}
\hline
& A & 2 & 3 & 5 & 13 & 29 & 39 & 107 & 503 & 1021 \\
N & M \\
\hline
1 & 2 & 0:0  & 0:0  & 0:0  & 0:0  & 0:0  & 0:0  & 0:0  & 0:0  & 0:0 \\
1 & 3 & 0:0  & 0:0  & 0:0  & 0:0  & 0:0  & 0:0  & 0:0  & 0:0  & 0:0 \\
1 & 4 & 0:0  & 0:0  & 0:0  & 0:0  & 0:0  & 0:0  & 0:0  & 0:0  & 0:0 \\
1 & 5 & 0:0  & 0:0  & 0:0  & 0:0  & 0:0  & 0:0  & 0:0  & 0:0  & 0:0 \\
1 & 6 & 0:0  & 0:0  & 0:0  & 0:0  & 0:0  & 0:0  & 0:0  & 0:0  & \\
1 & 7 & 0:0  & 0:0  & 0:0  & 0:0  & 0:0  & 0:0  & 0:0  & 0:0  & \\
1 & 8 & 0:0  & 0:0  & 0:0  & 0:0  & 0:0  & 0:0  & 0:0  &  & \\
1 & 9 & 0:0  & 0:0  & 0:0  & 0:0  & 0:0  & 0:0  &  &  & \\
\hline
2 & 2 & 0:0  & 0:0  & 0:0  & 0:0  & 0:0  & 0:0  & 0:0  & 0:0  & 0:0 \\
2 & 3 & 0:0  & 0:0  & 0:0  & 0:0  & 0:0  & 0:0  & 0:0  & 0:0  & 0:0 \\
2 & 4 & 0:0  & 0:0  & 0:0  & 0:0  & 0:0  & 0:0  & 0:0  & 1:0  & 0:0 \\
2 & 5 & 0:0  & 0:0  & 0:0  & 0:0  & 0:0  & 0:0  & 0:0  & 0:4  & \\
2 & 6 & 0:0  & 0:0  & 0:0  & 0:0  & 0:0  & 0:0  & 0:0  &  & \\
2 & 7 & 0:0  & 0:0  & 0:0  & 0:0  & 0:0  & 0:0  &  &  & \\
2 & 8 & 0:0  & 0:0  & 0:0  & 0:0  & 0:0  & 0:0  &  &  & \\
\hline
3 & 3 & 0:0  & 0:0  & 0:0  & 0:0  & 0:0  & 0:0  & 0:0  & 0:0  & \\
3 & 4 & 0:0  & 0:0  & 0:0  & 0:0  & 0:0  & 0:0  & 0:0  &  & \\
3 & 5 & 0:0  & 0:0  & 0:0  & 0:0  & 0:0  & 0:0  & 0:0  &  & \\
3 & 6 & 0:0  & 0:0  & 0:0  & 0:0  & 0:0  & 0:0  &  &  & \\
\hline
4 & 4 & 0:0  & 0:0  & 0:0  & 0:0  & 0:0  & 0:0  & 0:0  &  & \\
4 & 5 & 0:0  & 0:0  & 0:0  & 0:0  & 0:0  & 0:0  &  &  & \\

\hline
\end{tabular}
\caption{\label{fig3}Number of timeouts ($10$ minutes) for the Graph and Slopes algorithms on each class.}
\end{center}
\end{figure}

\begin{figure}
\begin{center}
\begin{tabular}{c@{\hspace{0.5cm}}c@{\hspace{0.5cm}}c@{\hspace{0.5cm}}c@{\hspace{0.5cm}}c@{\hspace{0.5cm}}c@{\hspace{0.5cm}}c@{\hspace{0.5cm}}c@{\hspace{0.5cm}}c@{\hspace{0.5cm}}c@{\hspace{0.5cm}}c}
\hline
& A & 13 & 29 & 39 & 107 & 503 & 1021 \\
N & M \\
\hline
1 & 2 & .0:.0  & .0:.0  & .0:.0  & .0:.0  & .0:.0  & .1:.0 \\
1 & 3 & .0:.0  & .0:.0  & .0:.0  & .0:.0  & .1:.0  & .1:.0 \\
1 & 4 & .0:.0  & .0:.0  & .0:.0  & .0:.0  & .2:.0  & .8:.0 \\
1 & 5 & .0:.0  & .0:.0  & .0:.0  & .1:.0  & 2.7:.1  & 5.7:.2 \\
1 & 6 & .0:.0  & .0:.0  & .0:.0  & .1:.0  & 9.8:.3  & \\
1 & 7 & .0:.0  & .0:.0  & .1:.0  & .4:.1  & 20.5:.4  & \\
1 & 8 & .0:.0  & .1:.1  & .1:.0  & 1.8:.5  &  & \\
1 & 9 & .0:.0  & .1:.1  & .2:.0  &  &  & \\
\hline
2 & 2 & .0:.0  & .0:.0  & .0:.0  & .0:.0  & .1:.0  & .4:.1 \\
2 & 3 & .0:.0  & .0:.0  & .0:.0  & .1:.1  & .2:.5  & 2.9:3.0 \\
2 & 4 & .0:.0  & .0:.0  & .1:.0  & .2:.3  & 602.1:190.0  & 16.8:550.0 \\
2 & 5 & .0:.0  & .0:.0  & .1:.1  & .2:.7  & 25.2:3333.7  & \\
2 & 6 & .1:.0  & .2:.2  & .1:.2  & 1.1:12.0  &  & \\
2 & 7 & .0:.0  & .1:.1  & .1:.1  &  &  & \\
2 & 8 & .1:.0  & .2:.4  & .4:1.1  &  &  & \\
\hline
3 & 3 & .0:.0  & .1:.0  & .1:.0  & .1:.1  & 17.0:68.5  & \\
3 & 4 & .0:.0  & .1:.0  & .1:.1  & .8:1.7  &  & \\
3 & 5 & .0:.0  & .1:.1  & .1:.1  & 3.3:28.2  &  & \\
3 & 6 & .1:.0  & .1:.1  & .4:.6  &  &  & \\
\hline
4 & 4 & .1:.0  & .1:.1  & .2:.2  & 3.3:85.8  &  & \\
4 & 5 & .1:.0  & .2:.2  & .4:.6  &  &  & \\

\hline
\end{tabular}
\caption{
  \label{fig4}The total time, expressed in seconds, spent by each
  algorithm (Graph and Slopes) on all tests for each class. Columns
  with A $ < 13$ are excluded because the values are very close to zero. }
\end{center}
\end{figure}

\end{document}